\documentclass[USenglish,twocolumn,14pt]{article}

\usepackage[utf8]{inputenc}%(only for the pdftex engine)
\usepackage[big]{dgruyter}
\usepackage{graphicx}
\usepackage{color}
\graphicspath{{images/}}
\usepackage{url}
\usepackage{xspace}
\usepackage{amsmath}	% Advanced maths commands
\usepackage{amssymb}	% Extra maths symbols
\usepackage[varg]{txfonts}

\newcommand{\hen}{\mbox{Hen\,2-428}\xspace}
\newcounter{Rco}
\newcommand{\Ionst}[1]{\setcounter{Rco}{#1}\Roman{Rco}}
\newcommand{\Ion}[2]{\mbox{#1\,{\scriptsize\Ionst{#2}}}} % ie.\Ion{Fe}{2}
\newcommand{\Ionw}[3]{\mbox{#1\,{\scriptsize\Ionst{#2}}~$\lambda\,#3$\,\AA}} % ie. \Ionw{He}{2}{4686}
\newcommand{\Ionww}[3]{\mbox{#1\,{\scriptsize\Ionst{#2}}~$\lambda\lambda\,#3$\,\AA}} % ie. \Ionww{He}{2}{4200, 4542, 5412}
\newcommand{\logg}{\mbox{$\log g$}\xspace}
\newcommand{\loggw}[1]{\mbox{$\log g\hspace{-0.5mm} =\hspace{-0.5mm}  #1$}} % ie. \loggw{5.25 \pm 0.07}
\newcommand{\kK}{\mbox{\rm kK}\xspace}
\newcommand{\Teff}{\mbox{$T_\mathrm{eff}$}\xspace}
\newcommand{\Teffw}[1]{\mbox{$\Teff\hspace{-0.5mm} =\hspace{-0.5mm} #1 \,\mathrm{kK}$}} % ie. \Teffw{53.0 \pm 6.9} 

\newcommand{\Msol}{$M_\odot$}

\begin{document}

\articletype{Research Article{\hfill}Open Access}

\author*[1]{N. L. Finch}

\author[2]{N. Reindl}
\author[3]{M. A. Barstow}
\author[4]{S. L. Casewell}
\author[5]{S. Geier}
\author[6]{M. M. Miller Bertolami}
\author[7]{S. Taubenberger}

\affil[1]{Department of Physics and Astronomy, University of Leicester, University Road, Leicester LE1 7RH, UK, E-mail: nlf7@le.ac.uk}
\affil[2]{Department of Physics and Astronomy, University of Leicester, University Road, Leicester LE1 7RH, UK}
\affil[3]{Department of Physics and Astronomy, University of Leicester, University Road, Leicester LE1 7RH, UK}
\affil[4]{Department of Physics and Astronomy, University of Leicester, University Road, Leicester LE1 7RH, UK}
\affil[5]{Institute for Astronomy and Astrophysics, Kepler Center for Astro and Particle Physics, Eberhard Karls University, Sand 1, D-72076 T\"{u}bingen, Germany}
\affil[6]{Instituto de Astrofísica de La Plata, UNLP-CONICET, La Plata, Buenos Aires 1900, Argentina}
\affil[7]{Max Planck Institut f\"{u}r Astrophysik, Karl-Schwarzschild-Str. 1, D-85748, Garching, Germany}
\affil[7]{European Southern Observatory, Karl-Schwarzschild-Str. 2, D-85748, Garching, Germany}

  \title{\huge Spectral analysis of the binary nucleus of the planetary nebula Hen\,2-428 \textendash{} first results
}

  \runningtitle{Spectral analysis of Hen\,2-428
}

  %\subtitle{...}

  \begin{abstract}
{Identifying progenitor systems for the double-degenerate scenario is crucial to check the reliability of type Ia supernovae as cosmological standard candles. \citet{Santander15} claimed that \hen has a double-degenerate core whose combined mass significantly exceeds the Chandrasekhar limit. Together with the short orbital period (4.2\,hours), the authors concluded that the system should merge within a Hubble time triggering a type Ia supernova event. \citet{Garcia-Berro16} explored alternative scenarios to explain the observational evidence, as the high mass conclusion is highly unlikely within predictions from stellar evolution theory. They conclude that the evidence supporting the supernova progenitor status of the system is premature. Here we present the first quantitative spectral analysis of \hen which allows us to derive the effective temperatures, surface gravities and helium abundance of the two CSPNe based on state-of-the-art, non-LTE model atmospheres. These results provide constrains for further studies of this particularly  interesting system.
}
\end{abstract}
  \keywords{subdwarf, planetary nebula, type Ia}
%  \classification[PACS]{}
 % \communicated{...}
 % \dedication{...}

  \journalname{Open Astronomy}
\DOI{DOI}
  \startpage{1}
  \received{..}
  \revised{..}
  \accepted{..}

  \journalyear{2017}
  \journalvolume{000}
  \journalissue{000}
 
\maketitle

\section{Introduction}

Type Ia supernovae (SN Ia) are the most important standard candles to measure the largest structures of our universe. Although there is a general consensus that only the thermonuclear explosion of a white dwarf can explain the observed features of those events, the nature of their progenitor systems still remains elusive. One of the progenitor scenarios discussed is the merger of two white dwarfs with a total mass exceeding the Chandrasekhar mass limit, the double-degenerate model \citep{Iben84, Webbink84}. Another progenitor scenario is the single-degenerate model, in which a white dwarf accretes material from a non-degenerate companion and explodes when it reaches the mass limit \citep{Whelan73}.

\citet{Santander15} discovered that the central star of the planetary nebula (CSPN) \hen is a close binary with an orbital period as short as 4.2\,hours based on sinusoidal variations in its light curve. Furthermore, they detected a time-dependent splitting of the \Ionw{He}{2}{5411} line, which they identified as a signature for a double-lined binary system consisting of two pre-white dwarfs with equal masses of 0.88\,\Msol. In this case, the system would merge within 700 million years and since its total mass would exceed the Chandrasekhar limit, it would be one of the best SN Ia progenitor candidates known. 

This scenario has since been challenged by \citet{Garcia-Berro16}. The most important criticism is the strong mismatch between the luminosities and radii of both white dwarf components as derived by \citet{Santander15} with the predictions from stellar evolution models \citep{Bloecker91}. Furthermore, the evolutionary times for post-AGB stars with masses as high as 0.88\,\Msol\,\,to join the white dwarf cooling curve are predicted to be of the order of hundreds of years or less \citep{Bertolami16}. Therefore, it should be almost impossible to observe two such objects in exactly the same stage.

\citet{Garcia-Berro16} also perform a photometric analysis, but used evolutionary tracks of \citet{Renedo10} to constrain the stellar temperatures. They showed that the light curves of \hen may also be fitted well by assuming two CSPNe with lower masses (i.e., masses of 0.47\,\Msol\,\,and 0.48\,\Msol) in an overcontact binary. Thus, \citet{Garcia-Berro16} conclude that the claim that \hen provides observational evidence for the double degenerate scenario for SN Ia is premature.

Given the potential importance of \hen as a unique laboratory to study the double degenerate merger scenario, it is highly desirable to resolve this debate. Here, we provide, for the first time, a spectroscopic analysis of the system in order to constrain the effective temperatures, surface gravities and helium abundance through state of the art non-LTE atmospheric models.

\begin{figure*}
\centering
\includegraphics[scale=0.50]{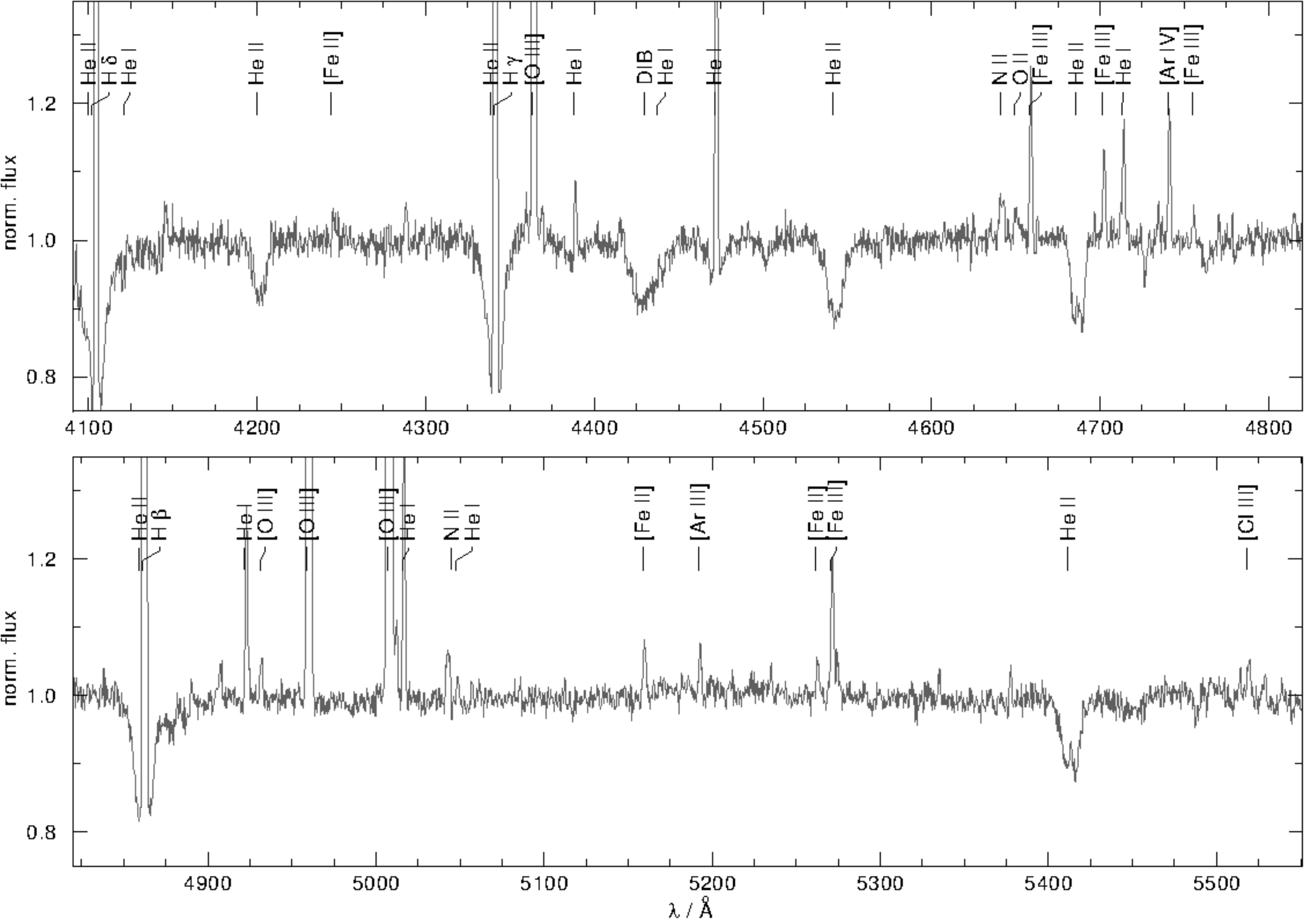}
\caption{FORS2 spectrum of \hen. Identified lines are marked.}
\label{fig:spectrum}
\end{figure*}

\section{Observations}

We downloaded the VLT/FORS2 spectra of \hen from the ESO archive (ProgIDs 085.D-0629(A) and 089.D-0453(A)). The observations were performed at different phases of the orbital period using the 1200G grism (resolving power $R=$\,$~1605$). The spectra were reduced using standard IRAF procedures.
%=lambda/delta lambda

%\textcolor{red}{list all nebula lines. Also check have neb lines been seen before in other papers? If not: For the first time we were able to identify nebular lines of...}
A FORS2/1200G spectrum of \hen is shown in Fig.~\ref{fig:spectrum}. We identified photospheric absorption lines of \Ion{H}{1}, \Ion{He}{1} and \Ion{He}{2}, as well as nebular emission lines of \Ion{H}{1}, \Ion{He}{1}, \Ion{N}{2}, \Ion{O}{2}, $[$\Ion{O}{3}$]$, $[$\Ion{Cl}{3}$]$, $[$\Ion{Ar}{3}$]$, $[$\Ion{Ar}{4}$]$, $[$\Ion{Fe}{2}$]$, and $[$\Ion{Fe}{3}$]$, and a diffuse interstellar band (DIB) at 4430\,\AA. The \Ion{H}{1} and \Ion{He}{1} lines are clearly contaminated by nebula emission. The \Ion{He}{2} absorption features are double lined and differ between spectra, but the cause of this variability is disputed. \citet{Santander15} explained this as a super-composition of two photospheric absorption lines, one from each star, varying due to orbital motion. However, \citet{Garcia-Berro16} speculated that the variability in the \Ionw{He}{2}{5411} line could be interpreted as a narrow \Ion{He}{2} emission line with varying strength that originates from the compact planetary nebula superimposed onto a broad absorption line from the central stars. On the other hand, the variability of the \Ionw{He}{2}{4686} line cannot be explained by the presence of a nebular emission line. Thus, we conclude that all four \Ion{He}{2} features are purely photospheric, double lined, and the variability is due to the orbital motion. Therefore, these lines are suitable to perform the spectral analysis on.

%% ###################################################################

\begin{figure*}
\centering
\includegraphics[scale=0.63]{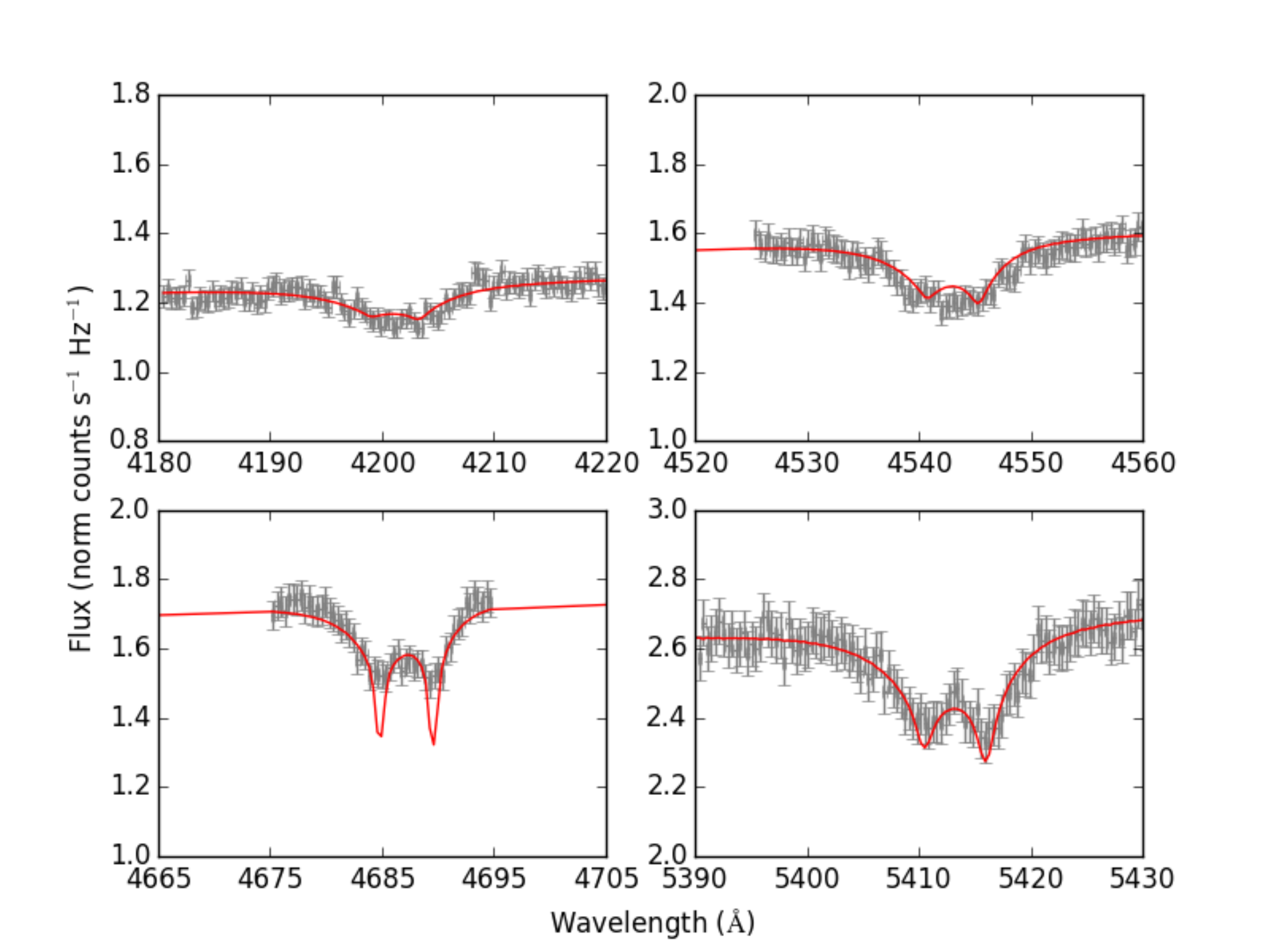}
\caption{\label{fig:fit} The model spectrum (red) plotted against the FORS2 spectrum (grey) of \hen for four \Ion{He}{2} lines. The model has stellar temperatures of 53.0\,\kK and 49.0\,\kK, and \logg of 5.25 and 4.91. Helium mass fractions are 0.27 and 0.24 respectively.}
%\label{fig:fit} 
\end{figure*}

%\section{Results}
\section{Spectral Analysis}

For the spectral analysis we employed the T\"{u}bingen Model Atmospheres Package (TMAP, \citet{Werner03}, \citet{TMAP}, \citet{Rauch03}) to produce state-of-the-art non-LTE model atmospheres. Initially, our models contained only hydrogen and helium. The model grid spans from \Teff\,$=$\,30.0 - 70.0\,\kK (2.5\,\kK steps), \loggw{4.25 - 6.00} (0.25 steps), and covered three helium abundances ($\log\,(He/H) = -2, -1, 0$, number fractions).

To constrain the parameters of the system we used the XSPEC software \citep{Shafer91, xspec}. XSPEC is a chi square minimisation code, originally designed for X-ray spectra, but which has been adapted to work on optical data \citep{Dobbie04}. XSPEC determines the best fit model for the input parameters, which in this case are the effective temperatures, surface gravities, helium abundances, and radial velocities. First the radial velocity of each component was found, and then these were fixed whilst deriving \Teff, \logg and $\log\,(He/H)$ simultaneously. The flux contribution from each star is accounted for in XSPEC by adding our model grid to itself, thus allowing us to treat each of the stars' parameters separately and, therefore, replicating the binarity of the system.

Our preliminary results give \Teffw{53.0 \pm 6.9}, \loggw{5.25 \pm 0.07} and a helium mass fraction of 0.27\,$\pm$\,0.05 for the first star. For the second star we obtained \Teffw{49.0 \pm 6.9}, \loggw{4.91 \pm 0.08} and a helium mass fraction of 0.24\,$\pm$\,0.05. The reduced chi squared value was 0.65 for these parameters. The results are displayed visually in Fig.~\ref{fig:fit}, where grey lines show the observed spectrum from one of the 2012 observations, and the red line is our best-fit TMAP model.

Whilst \Ionww{He}{2}{4200, 4542, 5412} are reproduced nicely, the line cores of \Ionw{He}{2}{4686} appear too deep compared to the observation. This is a known problem when fitting the spectra of CSPNe and other hot stars with pure hydrogen and helium models only. It has been shown that \Ionw{He}{2}{4686} is particularly susceptible to metal line blanketing \citep{Reindl14, Latour15}. Therefore, as a next step, we will also include metals in our TMAP models. \citet{Rodriguez01} measured the nebula abundances of O, N, S, and Ar of \hen, which will provide an estimate of photospheric metal abundances of the stars.
% and \Teff$_2$ =49000$ \pm$ 6900&\,K, and log$g$ values of 5.25\,$\pm$\,0.07 and 4.91\,$\pm$\,0.08 respectively. Helium mass fractions of 
%0.269\,$\pm$\,0.048 and 0.236\,$\pm$\,0.049 were also constrained. 
%Using hydrogen and helium models only, the 5412\,\AA, 4542\,\AA, and 4200\,\AA\,\,absorption lines are modelled well by the best fit results. However, in all spectra the 4686\,\AA\,\,lines are not comparable to the model, as seen in the bottom left panel of Fig. \ref{spectra}. 

%We also stress that the synthetic spectra show that temperatures above 33\,\kK are required to produce the \Ionww{He}{2}{4200, 4542, 5412} lines at the highest helium abundances of our grid, $\log\,(He/H) = 0$. However, this is also true at lower abundances, such as $\log\,(He/H) = -1$ which is close to the value derived for the CSPN of \hen. We note that higher temperatures are required if the gravity increases beyond that which was derived. The \Ionw{He}{2}{4686} line is visible at our lowest temperature, 30\,\kK. The lack of \Ionww{He}{2}{4200, 4542, 5412} lines in our lower temperature synthetic spectra is directly opposing the results from \citet{Santander15} who present a photometric model with stellar temperatures of 32.4\,$\pm$\,5.2\,\kK and 30.9\,$\pm$\,5.2\,\kK. No spectral analysis is given in their work.

Our derived effective temperatures (\Teffw{53.0\,\pm\,6.9}, \loggw{5.25\,\pm\,0.07} and \Teffw{49.0\,\pm\,6.9}, \loggw{4.91\,\pm\,0.08}) contradict the results from \citet{Santander15} who present a photometric model with stellar temperatures of 32.4\,$\pm$\,5.2\,\kK and 30.9\,$\pm$\,5.2\,\kK. In our synthetic spectra the \Ionw{He}{2}{4686} line is present even at our lowest temperature, 30\,\kK. However, we stress that the synthetic spectra show that temperatures above 33\,\kK are required to produce the \Ionww{He}{2}{4200, 4542, 5412} lines at the highest helium abundances of our grid, $\log\,(He/H) = 0$. This is also true at lower abundances, such as $\log\,(He/H) = -1$ which is close to the value derived for the CSPNe of \hen \citep{Rodriguez01}. For this helium abundance temperatures exceeding \Teff$\approx 40$\,kK are needed to reproduce \Ionww{He}{2}{4200, 4542, 5412} at their observed strengths. Even higher temperatures are required to produce these lines if the gravity is higher than our derived value. 

As already noted by \citet{Garcia-Berro16}, the upper limit placed by \citet{Santander15} on the effective temperatures of \Teff$\le 40$\,kK of the two CSPNe of \hen is not valid. \citet{Santander15} established this upper limit based on the absence of a \Ionw{He}{2}{5412} emission line, but there are many PNe whose central stars have effective temperatures larger than 40\,kK and do not have He II emission lines. In particular, \citet{Garcia-Berro16} showed that the light curves of \hen can be fitted equally well with models that assume 40-45\,\kK, which is consistent with our results.

\section{Future Work}

We have carried out the first quantitative spectral analysis of \hen and constrained the effective temperatures, surface gravities and helium abundance of the two CSPNe based on the VLT/FORS2 spectra. We will extend our spectral analysis to include the OSIRIS spectra taken at the Gran Telescopio Canarias to place stronger constrains on the atmospheric parameters of the CSPNe. This is the first step towards a more comprehensive analysis of this exceptional system. In a next step, we envisage a more sophisticated radial velocity analysis by measuring the radial velocities of the two components using our synthetic spectra, in contrast to the Gaussian fitting of \citet{Santander15}. In particular, we will take into account all four \Ion{He}{2} lines for these measurements instead of only \Ionw{He}{2}{5412} used by \citet{Santander15}. The derived radial velocities will provide more reliable dynamical masses, which are the key point in understanding the true nature of the system. The dynamical masses, as well as the effective temperatures derived by our spectral analysis, will then serve as steady base for the light curve models and allow a sound conclusion on the true nature of the CSPNe of \hen.

%\section{Conclusions}

%We carried out the first quantitative spectral analysis of \hen and constrained the effective temperatures, surface gravities and helium abundance of the two CSPNe. Comparing the preliminary results of this analysis to recent stellar evolutionary calculations we find that \hen likely consists of two post-RGB stars. This implies that the total mass of \hen is well below the Chandrasekhar limit and therefore the system will not trigger a SN Ia event.

\section*{Acknowledgements}

This work is based on observations collected at the European Organisation for Astronomical Research in the Southern Hemisphere under ESO programmes 085.D-0629(A) and 089.D-0453(A). NF is supported by an STFC Studentship. NR was supported by a Leverhulme Trust Research Project Grant. NR is supported by a research fellowship of the Royal Commission for the Exhibition of 1851. The TMAD service (\url{http://astro-uni-tuebingen.de/~TMAD}) used to compile atomic data for this work was constructed as part of the activities of the German Astrophysical Virtual Observatory. This research used the SPECTRE and ALICE High Performance Computing Facilities at the University of Leicester.

\end{document}